# The AI Scaling Wall of Diminishing Returns

*Of LLMs, Electric Dogs, and General Relativity*

Hemant Shukla

**Artificial Thinking**

My very first interaction with an LLM was with ChatGPT when it launched. I tested it on a physics problem I'd worked out decades ago regarding estimating the Sun's radiation in Earth's shadow at L2. After several iterations, nudging, and corrections, ChatGPT finally responded with a literal "Aha!" and claimed it *understood* what I was trying to accomplish and blurted out more than I had asked for.

It all felt eerily familiar to a scene from a short sci-fi story I once wrote (circa 2000-2001) called *Teaching Eter*. In it, a professor, while training Eter, an omnipresent, always-learning network that could access anything and everything, realizes that it has just learned something on its own. The name Eter was a play on the Rete algorithm I was using at the time while building a rule-based decision engine for the Hubble Space Telescope.

In that "aha" moment with ChatGPT, fiction seemed to morph into reality.

**The Two Worlds Of AI**

Suddenly, the world divided into two groups: those who construct the LLMs, and those who use them.

Since that first interaction, only three years ago, the models have grown larger and better. Technically, larger in terms of tokens and parameters, and better by the benchmarks they are tested against, such as Massive Multitask Language Understanding (MMLU; Hendrycks et al., 2020) and its newer variant, MMLU-Pro (Wang et al., 2024). These tests attempt to measure a model's general knowledge and reasoning across a wide range of tasks.

Meanwhile, in the other world, AI startups are springing up like mushrooms after an LLM rain[1]. The disruption to life as we've known it across industries is loud and clear.

---

[1] Over $450 billion has flowed into AI ventures globally since 2020, with the United States accounting for a significant majority of that total. In 2025 alone, AI startups are on track to draw a projected record of nearly $200 billion in new funding, representing over half of all global venture capital. Point to note, investment is highly concentrated in a few "frontier" and "foundational" model companies.

Despite the hesitation of the unknown, amid winds of caution and cries of doom, the underlying value seems immense. There's no stopping the AI rush.

There is hope that by simply building ever-bigger LLMs, we might reach the coveted state of artificial general intelligence (AGI) - that loosely defined notion of AI that can reason, learn, and adapt like humans.

**The Electric Dog Problem**

I'm not so sure, though. LLMs may seem to mimic a thought process, but cosine similarity across high dimensional embeddings of human information is a stretch to claim the same as thinking. I'm in the Roger Penrose camp: I don't think brains compute. There are no differential equations being solved in a dog's brain as it leaps to catch a frisbee - likely some deeper, biological mathematics we've yet to discover.

But that same dog, however naturally adept at catching a frisbee, will never learn general relativity. There's a limit to its learning setup.

Do LLMs face a similar intrinsic limit? That may be a trickier question to address. Perhaps a more reachable and relatively precise one would be: how do LLMs scale in the better-vs-large space?

**Measuring Intelligence**

The simplest way to answer the question was fitting a non-linear chi-squared against a polynomial to the models in some form of better (accuracy) vs large (compute size) domain. However, the problem with such an approach was that the fitted functions measured how the current models performed, and they consistently remained below the 90% threshold for AGI. There was no scaling behavior to be derived from such a fit.

Fortunately, Hoffmann et al. (*Hoffmann 2022*) had already laid the groundwork in their paper, *"Training Compute-Optimal Large Language Models."* Building on Kaplan et al. (2020), they showed model loss[2] follows a power-law with model size, dataset size, and compute.

To project Hoffmann's loss-scaling law into the accuracy–compute space, I selected MMLU-Pro as the benchmark for measuring performance against compute. Among the many available evaluations, MMLU-Pro provides the most transparent and self-reported dataset spanning both foundational and frontier models, making it uniquely suited for consistent cross-model comparison.

---

[2] The loss is defined as the average log likelihood of the correct next token. $L = -\frac{1}{N}\sum_{i=1}^{N} log\, p_\theta(x_i \mid x_{<i})$, where, $N$ is number of tokens, $x_i$ is the true next token, $x_{<i}$ is the preceding context, and $p_\theta$ is the model probability distribution assigned by the model with learned $\theta$ parameters.

As the broadest standardized assessment of general knowledge, reasoning, and instruction-following across disciplines, MMLU-Pro aligns closely with how models perform on real-world tasks, how humans rate their usefulness, and how well they apply learned knowledge to new problems, thus, making it a practical, if imperfect, proxy for evaluating emergent capability.

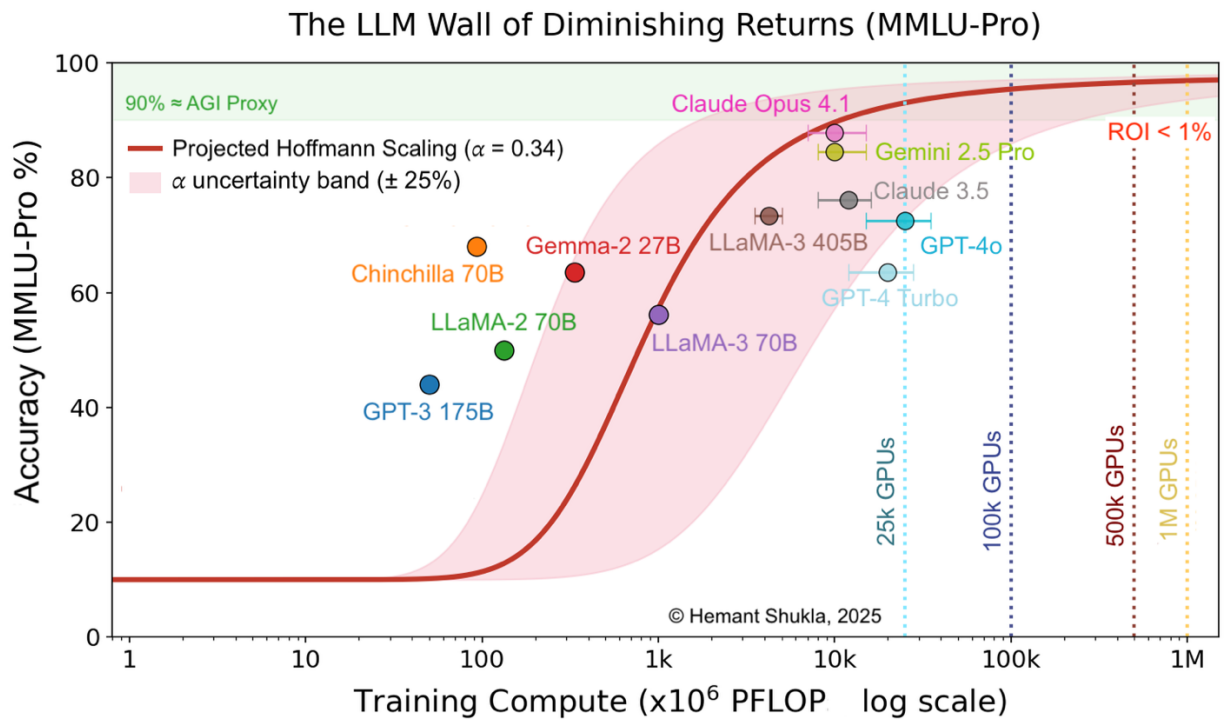

*Figure 1 - Scaling of MMLU-Pro accuracy vs. total training compute PFLOP. MMLU-Pro accuracy (%) of major foundational (< $10^{10}$ PFLOP) and frontier (≥ $10^{10}$ PFLOP) models plotted against total training compute. The fitted red curve and shaded uncertainty band ($\alpha = 0.34 \pm 0.08$) illustrate the predicted scaling behavior: rapid early gains that flatten beyond the YottaFLOP scale, approaching an asymptotic limit near 95 % accuracy.*

I defined a sigmoid function anchored to real data from Llama 3 70B (~1 billion PFLOP of total training compute) and Claude 4.1 (~10 billion PFLOP). The compute scaling exponent, $\alpha = 0.34$ from Hoffmann's formulation, was incorporated into the loss-to-accuracy transformation to retain the empirical dependence of loss on compute. A $\pm 25\%$ variation ($\alpha = 0.34 \pm 0.08$) was applied to generate the uncertainty band in the MMLU-Pro accuracy curve. While several other curve families were explored, including Gompertz, Richards, and Morgan-Mercer-Flodin, the sigmoid function provided the most stable and interpretable fit across anchors without violating the scaling constraints. The close convergence of all families at high-compute regimes further reinforced the robustness of the observed scaling behavior.

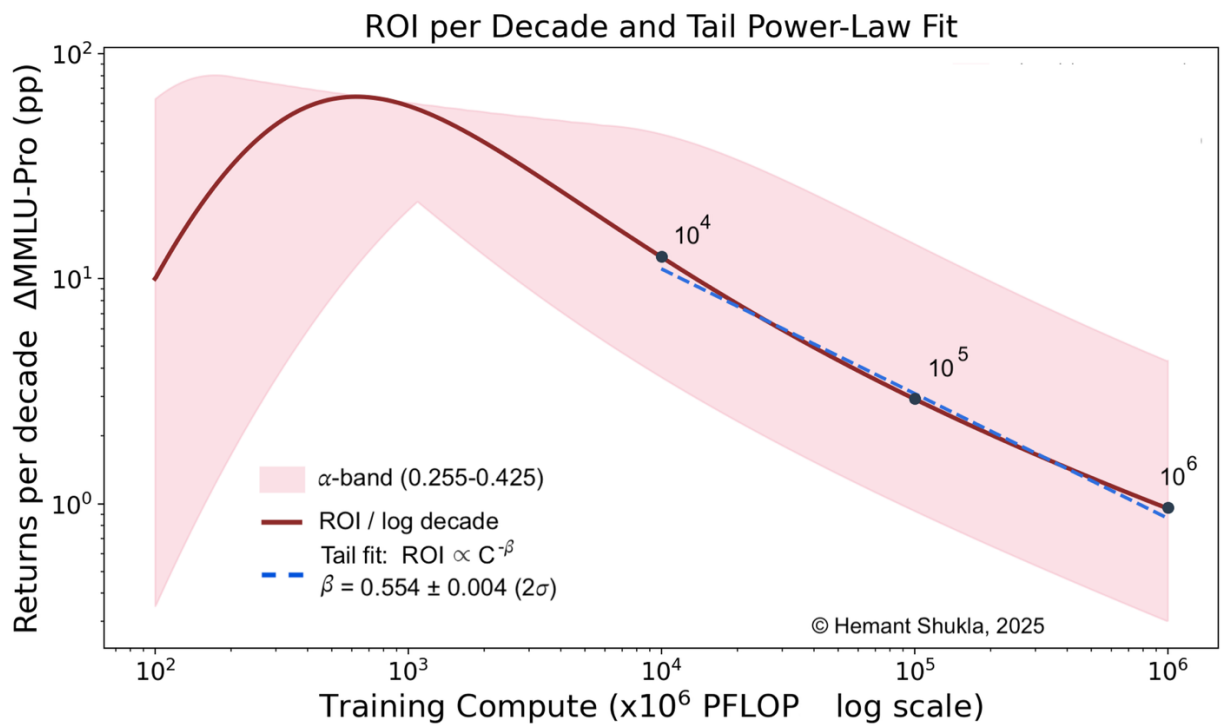

*Figure 2 -* Derivative of the scaling curve showing ROI fall-off. Derivative d*(Accuracy)/*d*(log₁₀ Compute)* quantifies the marginal MMLU-Pro accuracy gain per 10x increase in compute. The power-law fit gives *β* ≈ 0.55, indicating that each decade of additional compute yields only ≈ 1-2 percentage points (pp) of improvement.

**The Math Of Diminishing Returns**

The scaling relation built from the Hoffmann formulation is shown in Figure 1. The plot shows MMLU-Pro accuracy sharply rising and flattening beyond the yotta-scale. The fitted curve (red solid line) with ±25% *α*-band (pink) reaches the putative AGI-level threshold (90%) at roughly $10^{10}$ PFLOP of total compute and approaches an asymptote near 95%. Early models that sit above the curve likely benefited from architectural or data-quality innovations, whereas frontier models cluster along the predicted trend. For reference, we treat < $10^{10}$ PFLOP as *foundational* and ≥ $10^{10}$ PFLOP as *frontier*.

To quantify diminishing returns, Figure 2 plots the derivative $d(\text{Accuracy})/d(\log_{10} \text{Compute})$, i.e., the marginal accuracy gain per 10x increase in compute.

This relationship follows the empirical form,

$$\text{ROI pp/decade} \approx \Delta \text{ Accuracy per decade} \propto \text{Compute}^{-\beta}$$

A power-law fit to the tail yields $\beta = 0.554 \pm 0.004 (2\sigma)$, implying that each decade of additional compute buys only ≈ 1-2 percentage points (pp) of MMLU-Pro improvement.

**What Does It All Mean - Part I**

Does this mean I won't have my HAL, Data, or Number Six? I thought the days of staring at a barely-hanging-in-there browser, awaiting AI's response to my cleverly crafted prompt were long gone?

From a purely resource standpoint, the implications of the plots are sobering.

Reaching 90% (AGI proxy) MMLU-Pro requires ≈ $10^{10}$ PFLOP (~ 10 YottaFLOPs) of total training compute. Pushing to 95 % demands ≈ $5.6 \times 10^{10}$ PFLOP (~ 56 YottaFLOPs), over five times more compute for just five additional percentage points of accuracy. And I doubt that final 5% represents the evolutionary gap between a chimpanzee and a human. The returns decline by a factor of ≈ 3.6 with each 10x increase in compute: at 10 YottaFLOPs, ~ 12 pp per decade; at 100 YottaFLOPs, ~ 3 pp; and beyond 1000 YottaFLOPs, < 1 pp.

These numbers alone rearrange the strategic calculus across the ecosystem. From VCs to startups to enterprise labs to research groups, the message is unmistakable: *scale alone is no longer the path to AGI*.

And yet, it's a familiar story. Every graph that flattens signals the end of progress, right before the next axis gets invented.

**We've Been Here Before, Many Times**

Freshly fallen from AI hype, confused and dazed in the foggy valley of the real, I recall: I've seen this before. Many times. I call it *chasing the pendulum*.
In early 2007, while seeking ways to accelerate realistic simulations (now often termed digital twins), I discovered GPUs - then a niche technology for gaming and animation. A few tests convinced me that for massively parallel workloads, this was transformative. But convincing others was harder. AMD's Brook framework barely existed; NVIDIA traded between $20 and $35. The establishment was skeptical.

So I launched an international GPU computing center at Berkeley Lab. Seven years of incremental proof followed including building one of the first GPU-based HPC clusters. Skepticism gave way to adoption. GPUs became the backbone of modern high-performance computing. AI emerged serendipitously, a consequence of trailblazing.

Now replace *GPUs* in that story with *AI*, and you'll arrive at today.

The new pendulum: underestimating a paradigm (GPU) in 2007, swinging all the way to overestimating a single dimension of it (AI) in 2025.

Every technology hits a wall. The question isn't whether the wall exists, the mathematics proves it does. The question is what happens next. In 2007, hitting GPU memory limits forced innovation in distributed and hybrid computing. In 2025, hitting AI scaling limits will force innovation in efficiency, architecture, and deployment.

Constraints don't end progress. They redirect it.

**North Of The Scaling Wall**

Foundational models have shown that true innovation lies *north of the scaling wall*.

If I am a VC funding AI companies, this slow lane to AGI fundamentally alters the investment thesis. The billions that flowed into AI since 2020 were predicated on exponential returns. But we're now in the regime where marginal improvements cost exponentially more. Now I am suddenly asking, "where's my killer app?", not, "where is my bigger model?" Everyone had a website selling goods back during the dotcom boom. Taking a leaf from that era, I'm looking for my Amazon. I am searching for a startup whose value increases even if model accuracy improves by only 1-3 pp per decade. In other words, a business that compounds on deployment efficiency, proprietary workflows, and recurring usage. *Scale is no longer the differentiator; compounding usage is.*

If I'm a founder or CEO building an AI startup, the ground beneath me has shifted. Model makers, seeing slowing gains, are moving up the stack, adding application layers from code generation to filmmaking. To compete, I must bring a unique *PageRank* to an overcrowded search space. Find a niche. Serve an underrepresented market. *The world of model builders may slow down, but the other world of users just lit up.*

If I'm deploying AI in enterprise, I can no longer afford to wait for AGI. Current frontier models represent ~90% of maximum achievable performance on general benchmarks. The next generation will be marginally better at drastically higher cost. The productivity gains I'm seeking will come from better implementation of existing models. *Build now. Iterate on deployment, not on model selection.*

If I am a research lab, I am already lagging behind and cannot compete on training budgets. My competitive edge isn't "we'll train a better model", it is the deployment efficiency, domain specialization, user experience, proprietary data, or integration depth. The victors of this next phase won't be defined by their training budgets, but by the clever engineering that can bump models vertically without needing larger compute. The breakthroughs will come from: inference-time optimization (making models think), architectural efficiency (getting more from less), post-training methods (better alignment and instruction-following), and hybrid approaches (combining strengths of different paradigms). *The low-hanging fruit of pure scaling has been picked.*

© Hemant Shukla, 2025. Licensed under CC BY 4.0 (Creative Commons Attribution 4.0 International).
hshuklatmp@yahoo.com

If I am a leader of a nation, I am monitoring converging forces - technological, economic, social, and geopolitical - that are swirling into unprecedented disruption. I am revising labor policies to absorb workforce displacements already unfolding due to AI. When faced with rising power budgets for bigger AI, I ask difficult questions: have we truly exhausted what we already possess before chasing the next frontier? If "average" AI has forced us to rethink employment and ethics, what will "advanced" AI (AGI) demand of us? The last time we harnessed new tools without foresight, we burnt the sky (altered the climate), and have only just begun to pay the price. A century's progress turned into an existential debt. We must not repeat that mistake with AI or other emerging technologies. There's a group outside my office eagerly waiting to show off their quantum trinkets and seek funding. Let us thoughtfully design a better future before it designs us. *The future is bright; let's arrive in it safely, not crash and burn into it.*

And finally, if I'm an LLM vendor, I'm asking myself a simple question: why am I confining one of the most interactive technologies of our lifetime to a static HTML textbox? Sure, most of my resources are diverted to building larger models, but shortchanging the interface cannot be a winning strategy. It's like engineering a powerful Porsche only for grocery runs.

I am also asking: why are we not learning from users? With user consent, every interaction could begin building dynamic, evolving user profiles. Such profiles could power proactive hiring, custom education, personalized onboarding, and even intent-aware commerce, where your digital twin does the browsing and buying for you - my dream project since I saw The Matrix. Isn't dodging bullets while buying *one* and getting another for free the quintessential metaphor for holiday shopping? Let your digital twin deal with that.

Done correctly, user profiles don't just redefine productivity; they rewrite the architecture of the web itself. The browser becomes an intelligent interface, and every user session becomes a feedback loop of context, preference, and purpose. *The potential is unbound.*

In my experiments, I continue to see spiking neural nets, high-dimensional computing, and variants of reinforcement learning, as tools that facilitate building safe, adaptable, personalized, and energy-efficient AI interfaces. *AI has a future in personalization for all.*

At this point, scale alone delivers vanishing returns - an ever-costlier pursuit of diminishing insight. Are we building a super-expensive tinfoil hat for the dog, desperately hoping it might learn general relativity?

**What Does It All Mean - Part II**

So, it appears that there may be an intrinsic limit to scaling-only approaches to LLMs in the accuracy-compute space.

My father often says, "Knowledge comes, but wisdom lingers." To my childhood ears, it made no sense. But today, in the context of the Scaling Wall, it is prescient. We have built machines with meaningless data and context-free information, with zero wisdom.

The "aha" moment I had with ChatGPT wasn't thinking; it was retrieval. It was a mirror reflecting my own physics problem back at me, just with better cosine similarity.

And perhaps that is the point. In *Teaching Eter*, I imagined a network that could *think*. But in reality, we built something arguably more useful: a network that can *fetch*.

The teams that win the next decade won't be the ones trying to force the dog to derive general relativity. They will be the ones who realize that while the dog can't do the math, it can fetch the chalk, the blackboard, and the papers of every physicist who ever lived, instantly.

Perhaps the future holds a wise AGI - one that can rearrange our world for the better. But it is highly unlikely to arrive on the wings of large transformers. If we are building a brain (AI), it has to be energy-efficient, scale invariant, and operating on bio-math.

*The flat curve is the new starting line. Let's build with what we have.*

*Acknowledgment:* I thank Dr. Horst Simon for his valuable feedback on this note.